\documentclass[nofootinbib, prd,12pt,preprint, longbibliography,aps]{revtex4-1}
\usepackage{graphicx}
\usepackage{epstopdf}
\usepackage{amsmath}
\usepackage{amssymb}
\usepackage{amsfonts}
\usepackage{amsthm}
\usepackage[usenames]{color}
\usepackage{array}
\usepackage{verbatim}

\usepackage[
      colorlinks=true,
      linkcolor=blue,
      urlcolor=blue,
      filecolor=blue,
      citecolor=red,
      pdfstartview=FitV,
      pdftitle={},
      pdfauthor={},
      pdfsubject={},
      pdfkeywords={},
      pdfpagemode={},
      bookmarksopen=true
]{hyperref}

\usepackage{epsfig}

\usepackage{amsfonts,amsthm,amsmath,amssymb,graphicx}
\usepackage{hyperref}
\usepackage{amsmath}
\usepackage{url}
\usepackage{graphicx,color}

 \def\frac#1#2{{#1\over #2}}

 \def\CN{{\cal N}}

\def\be{\begin{equation}}
\def\ee{\end{equation}}
\def\ba{\begin{eqnarray}}
\def\ea{\end{eqnarray}}
\numberwithin{equation}{section}

\begin{document}

\title{ Memory effect and BMS symmetries for extreme black holes }
\author{Srijit Bhattacharjee\footnote{srijuster@gmail.com} and Shailesh Kumar\footnote{shaileshkumar.1770@gmail.com }  }
\affiliation{Indian Institute of Information Technology, Allahabad\\
Devghat, Jhalwa, Uttar Pradesh-211015, India}

\begin{abstract}
We study horizon shells and soldering freedom for extreme black holes and how supertranslation-like Bondi-Metzner-Sachs (BMS) symmetries appear as soldering transformations. Further, for a null shell placed infinitesimally close to the horizon of an extreme Reissner-Nordstr$\ddot{o}$m (RN) black hole, we show superrotation-like symmetries also arise as soldering freedom. Next, considering the interaction of impulsive gravitational waves supported at the horizon shell with test particles, we study how the ``memory'' (or the imprints) of BMS-like symmetries gets encoded in the geodesics (test particles) crossing the shell. Our study shows, timelike test particles get displaced from their initial plane when they cross the horizon shell. For a null geodesic congruence crossing the horizon shell, the optical tensors corresponding to the congruence suffer jumps. In both the cases, the changes are induced by BMS parameters that constitute the gravity wave and matter degrees of freedom of the shell. \end{abstract}

\maketitle


\section{Introduction}
Gravitational memory effect has been an active area of research and it has attracted researchers across the disciplines- from classical gravity, gravitational wave astronomy \cite{Zeldovich:1974gvh, Braginsky-Thorne, *Braginsky, PhysRevLett.67.1486, Bieri:2013ada, Bieri1, PhysRevD.80.024002} to researchers in the area of quantum gravity \cite{Strominger:2014pwa, Strominger:2017zoo, PhysRevLett.116.231301, Zhang:2017geq, ZHANG2017743, Zhang:2017jma}. The feature that characterizes memory effect is a permanent displacement of test particles after the passage of a burst of gravitational waves. Memory effect may prove to be a very promising field where many predictions of classical General Relativity (GR) can be tested in the coming advanced detectors like advanced LIGO or LISA \cite{Islo:2019qht, PhysRevLett.117.061102}. On the other hand, its connection with soft theorems and asymptotic symmetries of spacetimes has given an intriguing chance to dig into the quantum structure of gravity in the low energy or infrared limit \cite{Strominger:2014pwa, Strominger:2017zoo, PhysRevLett.116.231301, Hawking:2016sgy}.

Extreme black holes are important to the GR, and String theory community, primarily because of its relevance in the calculation of black hole entropy \cite{Strominger:1996sh}. Extreme black holes allow conformal symmetry when we zoom in to its near horizon geometry, and this played a crucial role not only in the study of quantum black holes but also in finding new kinds of hairs in extreme black holes \cite{Angelopoulos:2018yvt}. As it has been studied that almost 70\% of astrophysical black holes are near extremal and many super-massive black holes are also near extremal \cite{Volonteri_2005, Gou_2014, McClintock_2006, Brenneman_2006, Brenneman_2011}, more attention needs to be paid in exploring the properties of these black holes.

In earlier studies \cite{Blau:2015nee, *Blau:2016juv, Bhattacharjee}, in the context of stitching of two spacetimes across a null hypersurface, it has been shown that BMS-like \cite{Bondi:1962px, *Sachs} asymptotic symmetries can be recovered at the event horizon of black holes when one demands the induced metric on the horizon remains invariant under arbitrary coordinate transformations owing to satisfy the junction conditions. As gluing two geometries across a null hypersurface (like event horizon of a black hole) generally produces thin shells containing some matter, these shells at the event horizon are termed as ``horizon shells'' \cite{PhysRevD.43.1129, C_Barrabes}. These horizon shells become a history of an Impulsive Gravitational Wave supported at the horizon. Similar kinds of studies in the context of plane-gravitational waves can be found in \cite{OLoughlin:2018ebk, ZHANG2017743, Zhang:2017jma}. Impulsive gravity waves are generated during violent astrophysical phenomena such as supernovae explosion, merger of heavy black holes etc. Therefore, it would be interesting to study the measurable effects of such signals on test detectors. In this note, we explore horizon shell in extreme black holes and the ``memory'' encoded in the geodesics crossing the shell. We find the BMS-like asymptotic symmetries are recovered at the horizon of the Extreme Reissner-Nordstr$\ddot{o}$m (ERN) black hole. We also find the shell-intrinsic quantities in different situations. Although astrophysical black holes are extremal in their rotation parameters, but due to lack of spherical symmetry, horizon shells in rotating spacetimes are difficult to analyze. For this, as a first step towards attempting an analysis of ``memory effect'' in the context of horizon shells in 4-dimensional rotating spacetime, we have considered an extreme RN black hole. This study can provide us some insights about more generic considerations. 

The recent extended version of BMS symmetry also contains superrotation symmetries which are defined as arbitrary conformal transformations on the celestial sphere \cite{PhysRevLett.105.111103}. In a shell placed close to horizon of an ERN black hole, it is shown how superrotation-type symmetries in the form of conformal transformations can be recovered near the horizon of ERN. We also briefly discuss the soldering symmetries of an extreme rotating Ba$\tilde{\text{n}}$adose-Teitelboim-Zanelli (BTZ) black hole. 

Next, we study the interaction of impulsive gravitational waves (IGW) with test geodesics crossing the horizon shell \cite{Penrose}. For a non-rotating and neutral black hole, it has been shown in \cite {PhysRevD.100.084010}, how BMS-like symmetries are encoded in the deviation vectors  for timelike geodesics crossing the horizon shells. Therefore, these geodesics or test particles carry imprints of BMS-like symmetries upon crossing the horizon shells and this is regarded as a ``memory effect''. Here, we repeat the analysis for the case of extreme black holes and show the memory is parametrized by BMS supertranslation parameters. For null geodesics, we see the effect of crossing a horizon shell supporting IGW is to trigger a discontinuity
in the $B$-tensor, whose different irreducible parts are the optical tensors (eg. expansion, shear etc.) \cite{PhysRevD.100.084010}. This jump in $B$ tensor components induced by the interaction of test geodesics with the shell is termed as ``$B$-memory'' \cite{OLoughlin:2018ebk}. For a horizon shell in an ERN black hole, we find that the effect of the passing of timelike geodesics across the horizon shell is to deflect the test particles off the initial surface where they were placed before the interaction of IGW. The displacements are determined by shell's intrinsic quantities that are parametrized by BMS transformation parameters. For null geodesics, we find jumps in the expansion and shear corresponding to the geodesic vector crossing the horizon shell transversely. Here again, the jumps are expressed in terms of BMS-like parameters. 

Let us briefly summarize how the draft has been organized. In section (\ref{sec2}), we briefly review the formulation based on which we would calculate the shell-intrinsic properties. In section (\ref{sec3}), we study the horizon shell in ERN spacetime and calculate the intrinsic properties of the shell. The connection of ERN shell with conformal symmetries is elucidated via studying shells close to the near horizon of an ERN metric. The horizon shells of extreme BTZ black holes are also discussed and the details are displayed in the appendix A. Section (\ref{sec4}) contains the description of memory effect for timelike detectors crossing an ERN horizon shell. In section (\ref{mNull}), introducing a B-tensor at the horizon and off-the horizon shell, $B$-memory effects for null detectors are discussed. Finally, we conclude with a discussion on the results of this study and indicate some future extensions of this work.  

\section{Horizon shells, IGW and BMS-like soldering freedom} \label{sec2}

The standard formalism of IGW relies on patching of two spacetimes. When two spacetimes are being glued across a null surface, the Riemann tensor expression produces a singular term proportional to the Dirac delta function. As a result, to satisfy Einstein field equation, the stress-energy tensor should also contain a singular piece associated with the matter present on the hypersurface. This is referred as a thin shell or a surface layer on the hypersurface. When the null hypersurface across which two spacetimes are glued becomes the event horizon of a black hole, the thin shell is termed as horizon shell. 

 Let us briefly discuss how IGW arise in the context of gluing two manifolds $\mathcal{M_{+}}$ and $\mathcal{M_{-}}$ across a common null hypersurface $\Sigma$. We shall use lower case Latin letters to denote hypersurface coordinates and Greek letters for spacetime indices. In other words, for 4-D case,  a spacetime index runs as $\mu = 0, 1, 2, 3$ and a hypersurface index runs as $a = 1, 2, 3$. Latin upper case letters are used to specify the (co-dimension one surface) spatial coordinates on the hypersurfaces. Let us setup the coordinates for $\mathcal{M_{\pm}}$ as $x^{\mu}_{\pm}$ with metrics $g^{\pm}_{\mu\nu}(x_{\pm}^{\mu})$ and a common coordinate system $x^{\mu}$ across the null hypersurface $\Sigma$. We denote $y^{a}$ as coordinates on the hypersurface with tangent vectors $e^{\mu}_{a}=\frac{\partial x^{\mu}}{\partial y^{a}}$. If we consider Kruskal type coordinates, $y^{a}$ can be written as $y^{a}=(V,y^{A})$ and we can write $e^{\mu}_{A}=\frac{\partial x^{\mu}}{\partial y^{A}}$.  The hypersurface is defined as $\Phi(x)=0$. Define the normal vector $n^{\mu}$ to the hypersurface as $n^{\mu}=g^{\mu\nu}\partial_{\nu}\Phi(x)$. We also define an auxiliary vector (non unique) $N^{\mu}$, which is transverse to the hypersurface $\Sigma$, with normalization conditions $N\cdot N=q$, $n\cdot N=-1$ and $e_{A}\cdot N=0$, to study the extrinsic properties of the shell. For convenience, $q$ is taken to be $0$  and $-1$ for $N^{\mu}$ null and time-like respectively. In terms of common coordinate system, we have 
 \[ [n^{\mu}]=[e^{\mu}_{a}]=[N^{\mu}]=0.\]
 Here $'[ \hspace{1mm} ]'$ denotes the difference of quantities between $+$ and $-$ sides. We must also mention here that a pseudo-inverse (non unique) of the degenerate metric $g_{ab}$ can be introduced \cite{PhysRevD.43.1129} and be denoted by $g_*^{ab}$. For studying extrinsic properties of the null hypersurface one can use a basis by pairing the four vectors $(N^{\mu}, e^{\mu}_a)$. In terms of these basis vectors one can write the inverse metric (or the completeness relation) as:
\begin{align}
g^{\mu\nu} = g^{ab}_{*}e^{\mu}_ae^{\nu}_b\,-n^{a}e_a^{\mu}N^{\nu }-n^{a} e_a^{\nu}N^{\mu }\label{c1},
\end{align}
with the condition 
\begin{align}\label{c2}
    g^{ab}_{*}N_{b}-n^{a}(N\cdot N)=0 .
\end{align} \cite{C_Barrabes}
The metric\footnote{Note that $N_{\mu} e^{\mu}_{a}=N_{a}$} takes the following form in this common coordinate system \cite{PhysRevD.43.1129, Poisson:2002nv}
\begin{align}
g_{\mu\nu} = g^{+}_{\mu\nu}\mathcal{H}(\Phi)+g^{-}_{\mu\nu}\mathcal{H}(-\Phi),
\end{align}
where $\mathcal{H}(\Phi)$ is Heaviside step function. And the junction condition is
\begin{align}
[g_{ab}] = g^{+}_{ab}-g^{-}_{ab}=0.
\end{align}
Using junction condition, we have
\begin{align}\label{riem3}
R^{\mu}{}_{\nu\rho\sigma} = R^{+\mu}{}_{\nu\rho\sigma}\mathcal{H}(\Phi)+R^{-\mu}{}_{\nu\rho\sigma}\mathcal{H}(-\Phi)+Q^{\mu}{}_{\nu\rho\sigma}\delta(\Phi),
\end{align}
where any non-zero component of the last term, given by $Q^{\mu}{}_{\nu\rho\sigma}=-\Big([\Gamma^{\mu}{}_{\nu\sigma}]n_{\rho}-[\Gamma^{\mu}{}_{\nu\rho}]n_{\sigma}\Big)$, implicates the existence of IGW on the null hypersurface. The full stress-energy tensor is given by
\begin{align}\label{3T}
T_{\mu\nu} = T^{+}_{\mu\nu}\mathcal{H}(\Phi)+T^{-}_{\mu\nu}\mathcal{H}(-\Phi)+S_{\mu\nu}\delta(\Phi),
\end{align}
where the stress-energy tensor $S^{\mu\nu}$ can be written as,
\begin{align}
S^{\mu\nu} = \mu n^{\mu}n^{\nu}+J^{A}(n^{\mu}e^{\nu}_{A}+n^{\nu}e^{\mu}_{A})+p\sigma^{AB}e^{\mu}_{A}e^{\nu}_{B}.
\label{ST}\end{align} 
The $S_{\mu\nu}$ projected on $\Sigma$ is, $S_{ab}=e^{\mu}_{a}e^{\nu}_{b}S_{\mu\nu}$. Where $\sigma^{AB}$ is non-degenerate metric for the spatial slice of the surface of the null shell; $A$ and $B$ denote the spatial indices. It is easy to deduce the singular part of the Einstein equation from (\ref{riem3}) and (\ref{3T}) in terms of $Q$-tensor and $S_{\mu\nu}$. Consequently, all measurable quantities on the shell are expressed in terms of the components of the stress-energy tensor. The intrinsic quantities of the shell are surface energy density, surface current and pressure denoted as $\mu$, $J^{A}$ and $p$ respectively, have the following form \cite{C_Barrabes, doi:10.1142/S0218271801001098, PhysRevD.43.1129, Poisson:2002nv}
\begin{align}
\mu = -\frac{1}{8\pi}\sigma^{AB}[\mathcal{K}_{AB}] \hspace*{4mm} ; \hspace*{4mm} J^{A} = \frac{1}{8\pi}\sigma^{AB}[\mathcal{K}_{VB}] \hspace*{4mm} ; \hspace*{4mm} p = -\frac{1}{8\pi}[\mathcal{K}_{VV}].
\end{align}
The properties of the shell are stored in $[\partial_{\alpha}g_{\mu\nu}]=\gamma_{\mu\nu}n_{\alpha}$ which says that the jump in the partial derivative of the metric is proportional to $\gamma_{\mu\nu}$. Furthermore, extrinsic curvature is related to the first derivative of $\gamma_{ab}$ on the shell in following way
\begin{align}
\gamma_{ab} = N^{\mu}[\partial_{\mu}g_{ab}]=2[\mathcal{K}_{ab}] \label{gamma from Kab},
\end{align}
together with
\begin{align}
\mathcal{K}_{ab}=e^{\mu}_{a}e^{\nu}_{b}\nabla_{\mu}N_{\nu} \label{curvature}.
\end{align}
$\gamma_{ab}$ is related to transverse traceless components of $\hat{\gamma}_{ab}$ by following relation
\begin{align}
\hat{\gamma}_{ab} = \gamma_{ab}-\frac{1}{2}g^{cd}_*\gamma_{cd}g_{ab}+2\gamma_{(a}N_{b)}+\Big(N_{a}N_{b}-\frac{1}{2}N.Ng_{ab}\Big)\gamma^{\dagger} \label{gamma cap},
\end{align}
where, $\gamma^{\dagger}=\gamma_{ab}n^{a}n^{b}$, $\gamma_{a}=\gamma_{ab}n^{b}$, and we also define $\gamma^{*}=\sigma^{AB}\gamma_{AB}$.


Now let us touch upon how BMS-like transformations occur when we glue two spacetimes across $\Sigma$. The details of the same can be seen in \cite{Blau:2015nee, *Blau:2016juv, Bhattacharjee, PhysRevD.100.084010}. BMS symmetries arise as the freedom of allowing general coordinate transformations (diffeomorphisms) that preserve the induced metric on the event horizon, when we patch two spacetimes across it. These possible transformations thus can be calculated by solving the Killing equation on the hypersurface \cite{Blau:2015nee, *Blau:2016juv}. So we require
\begin{align}
\mathcal{L}_{Z}g_{ab} = 0,
\end{align}
expanding this equation
\begin{align}
    Z^{c}\partial_{c}g_{ab}+(\partial_{a}Z^{c})g_{cb}+(\partial_{b}Z^{c})g_{ca} = 0\label{sup},
\end{align}
and working with $g_{aV}=0$ \footnote{We have adopted a Kruskal type coordinate system here.}:
\begin{align}
Z^{V}\partial_V g_{AB}+ Z^C\partial_C g_{AB}+\partial_A Z^C g_{BC}+\partial_B Z^C g_{AC}= 0 \label{next}.
\end{align}
Considering the metric $g_{AB}$ to be independent of $V$, then $Z^{V}$ remains unconstrained and can be of the form $Z^{V}=F(V,\theta,\phi)$. Further, if we set $\mathcal{L}_{Z}n^{a}=0$, one gets a restriction on $Z^{V}$
\begin{align}
\partial_{V}Z^{V} = 0 \hspace*{3mm} \Rightarrow \hspace*{3mm} Z^{V} = T(x^{A}).
\end{align}
It generates supertranslation-like transformation which can be written as
\begin{align}
V \rightarrow V+T(x^{A}).
\end{align}
\subsection{Superrotation}\label{supr}
 Recently an extended version of BMS symmetries has been obtained that contains a new kind of symmetry called ``superrotation'' at the asymptotic null infinities as well as near the horizon of black hole spacetimes \cite{PhysRevLett.105.111103, PhysRevLett.116.091101, Blau:2015nee, Bhattacharjee}. These are local conformal transformations of the celestial sphere at null infinities \cite{Barnich:2011ct, Strominger:2017zoo}. In \cite{Bhattacharjee}, it was also shown, if one relaxes the condition that $g_{AB}$ is independent of $V$, one can recover superrotation-like symmetries. This can be seen from Eq. (\ref{next}). If one allows conformal transformations to the spatial metric $g_{AB}$ and compensates that by a corresponding shift in $V$ direction, we can still satisfy  Eq. (\ref{next}) or the junction condition.  If a null hypersurface has a topology $\mathbb{R}\times S^2$, then one can allow conformal transformations on the (unit) sphere.  This reduces the part that contains purely spatial derivatives of Eq. (\ref{next}) to $\Omega(x^A)g_{AB}$\footnote{Using the relation $\mathcal{L}_Z g_{AB}=\Omega(x^A)g_{AB} $ for spatial components of $Z$. }. This means, one needs to find a solution of the following equation
\begin{align}\label{CKE}
    Z^{V}\partial_{V}g_{AB}+\Omega(x^{A})g_{AB} = 0.
\end{align}
In \cite{Bhattacharjee}, it has been shown that a solution can always be found if $g_{AB}$ can be expressed as a product of smooth functions of $V$, and $x^A$. Example of such cases are null cones of constant curvature spaces,  a null surface close to horizon of a black hole \cite{Bhattacharjee}.  The local conformal transformations which are being employed on $g_{AB}$ are thus equivalent to superrotations.  We shall discuss superrotations for extreme black holes in (\ref{supr1}).
\section{Intrinsic properties of Extreme RN black hole}\label{sec3}

In this section, we follow the recipe indicated in \cite{Bhattacharjee, PhysRevD.100.084010}. We have a seed ERN metric which is identified as manifold $\mathcal{M}_{-}$ and consider a metric for $\mathcal{M}_{+}$ manifold on which we do a supertranslation type coordinate transformation. The ERN metric in Eddington-Finkelstein (EF) coordinates for $\mathcal{M}_{-}$ manifold is given by
\begin{equation}
ds^{2} = -\Big(1-\frac{M}{r}\Big)^{2}dv^{2}+2dv dr+r^{2}(d\theta^{2}+\\sin^{2}\theta)d\phi^{2} \label{RN},
\end{equation}
with $v=t + r_*=t+\int \frac{dr}{(1-M/r)^2}$.\\

We perform the supertranslation type transformation on $v_+$ coordinate and keep other coordinates unaltered
\begin{align}
v_{+}=v+T(\theta,\phi) \hspace*{3mm} ; \hspace*{3mm} r_{+} = r \hspace*{3mm} ; \hspace*{3mm} \theta_{+}=\theta \hspace*{3mm} ; \hspace*{3mm} \phi_{+}=\phi. \label{supertranslation}
\end{align}

Under these transformations, the metric takes the following form
\begin{align}
ds_{+}^{2} =& -\Big(1-\frac{M}{r}\Big)^{2}(dv+T_{\theta}(\theta ,\phi)d\theta + T_{\phi}(\theta ,\phi)d\phi)^{2} \nonumber \\
& + 2(dv+T_{\theta}(\theta ,\phi)d\theta + T_{\phi}(\theta ,\phi)d\phi)dr+r^{2}d\Omega_{2}^{2},
\end{align} 
where the horizon is situated at $r=M$ and $T_{\theta}(\theta,\phi)=\partial_{\theta}T(\theta.\phi)$, $T_{\phi}(\theta,\phi)=\partial_{\phi}T(\theta.\phi)$. We choose the transversal (or the auxiliary null normal) vector $N_{v}=-1$. We use (\ref{curvature}) to compute the extrinsic curvature $\mathcal{K}_{\theta\theta}$, $\mathcal{K}_{\phi\phi}$ and $\mathcal{K}_{\theta\phi}$ on the horizon. Finally using (\ref{gamma from Kab}), we get expressions for $\gamma_{ab}$,
\begin{itemize}
\item[(i)] $\gamma_{\theta\theta} = 2\nabla_{\theta}T_{\theta}(\theta,\phi) $
\item[(ii)] $\gamma_{\phi\phi} = 2\nabla_{\phi}T_{\phi}(\theta,\phi)$
\item[(iii)] $\gamma_{\theta\phi} = \gamma_{\phi\theta} = 2\nabla_{\theta}T_{\phi}(\theta,\phi).$
\end{itemize}
Collectively, one can write $\gamma_{AB}=2\nabla_{A}T_{B}$. $\gamma_{AB}$ is symmetric in $A, B$. Notation $'\nabla'$ denotes the covariant derivative of $T(x^{A})$ with respect to the unit 2-sphere metric.  Now using (\ref{gamma cap}), one can also compute transverse traceless part of $\gamma_{ab}$
\begin{itemize}
\item[(i)] $\hat{\gamma}_{\theta\theta}  = \nabla_{\theta}T_{\theta}(\theta,\phi)-\frac{1}{\sin^{2}\theta}\nabla_{\phi}T_{\phi}(\theta,\phi) $
\item[(ii)] $\hat{\gamma}_{\phi\phi}  = \nabla_{\phi}T_{\phi}(\theta,\phi)-\sin^{2}\theta \nabla_{\theta}T_{\theta}(\theta,\phi)$
\item[(iii)] $\hat{\gamma}_{\theta\phi} = \hat{\gamma}_{\phi\theta} = 2\nabla_{\theta}T_{\phi}(\theta,\phi). $
\end{itemize}

\subsection{Shell-Intrinsic Properties}
With the help of extrinsic curvature expressions, we directly compute the surface energy density of the shell as,
\begin{align}
\mu = -\frac{1}{8M^{2}\pi}\bigtriangleup^{(2)}T(\theta,\phi).
\label{edc}\end{align}
The surface current $J^{A}$ and pressure of the shell is given by,
\begin{align}
J^{A} = 0 \hspace*{4mm} ; \hspace*{4mm} p = 0.
\end{align}
Interestingly, the current turns out to be zero in EF coordinates unlike the Schwarzchild or non-extreme case \cite{Blau:2015nee, *Blau:2016juv}. It is easy to see from (\ref{edc}),  the energy density is conserved 
along the null direction of the shell i.e. $\partial_v \mu=0$. However, there is no non-zero charge corresponding to this supertranslation-like BMS translation, as it vanishes when evaluated on the spherical surface.

Now let us try to see if we can get a shell without matter supporting only gravitational waves. For this, one needs to see if there is any regular solution of the equation obtained by setting $\mu$ equal to zero. 
\begin{align}
\bigtriangleup^{(2)}T(\theta,\phi) = 0.
\end{align} 
This is a Laplace's equation on a sphere. We know this Laplacian has spherical harmonics $Y_l^m(\theta, \phi)$ as eigenfunctions with $-l(l+1)$ as eigenvalues. However, here we have only $l=0$ as a feasible solution which corresponds to a constant only. Therefore, the allowed shift in the $v$ direction is of the form-
\be v\to v+ c,\ee
where $c$ is a constant. Direct substitution of this into the components of $\hat{\gamma}_{ab}$ yields zero. So there can't be a shell supporting pure gravitational waves. A similar situation was also obtained for Schwarzschild case \cite{Blau:2015nee, *Blau:2016juv}. Next, we would like to see if conformal symmetries can be recovered as soldering freedom in ERN spacetime. To see this, one must remember that shells in constant curvature spacetimes can produce conformal isometries as soldering freedom when Penrose's cut-and-paste approach is employed \cite{Penrose, C_Barrabes}. The basic technique is to glue two metrics across a null surface after performing a conformal transformation to the spatial part of the metric and a subsequent shift in the null direction on one side before attaching the other side \cite{ Penrose, pair creation}. In the following subsection, we wish to find conformal symmetries in ERN horizon shell from a similar kind of construction.
\subsection{Superrotation near the horizon of  ERN spacetime}\label{supr1}
As outlined in (\ref{supr}), superrotation-like soldering freedom can be obtained if one finds a metric that contains a $V$ dependent spatial part ($g_{AB}$) \cite{Bhattacharjee} at some $U$. Let us now examine this for a horizon shell in ERN spacetime. We introduce a null coordinate $U$ with the help of retarded null coordinate $u=t - r_*$, and write (\ref{RN}) as \cite{lucietti_horizon_2013}
\begin{equation}\label{ern-dn}
ds^2=-\frac {2f(r)}{f(M-U)}dUdv + r^2(U,v)d\Omega_2^2,
\end{equation}
where, 
\begin{equation}
f(r)=\left( 1-\frac{M}{r}\right)^2;\,\,r_*=r-M +2M\Big(ln\Big\vert\frac{r}{M}-1\Big\vert-\frac{M}{2(r-M)}\Big);\,u=-2r_*(M-U),
\end{equation}
and
\begin{equation}
2dU=\frac{du}{f(M-U)}.
\end{equation}
Using $r_*=(v-u)/2$ for small $U$ one finds $r(U,v)$ as
\begin{equation}\label{r-eqn}
r(U,v)=M-U+\frac{v}{2M^2}U^2 +\mathcal{O}(U^3)
\end{equation}

The metric (\ref{ern-dn}) is analytic at the event horizon where $U=0$. Now let us consider a null surface just outside the horizon.  For small $U\neq0$ the spatial section of the metric should now become a function of $v$. Therefore
as discussed in (\ref{supr}), we can recover superrotation-like symmetries on the shell situated close to horizon. To understand this clearly, we reparametrize the metric of unit sphere in terms of complex coordinates $z,\bar{z}$ and write
$d\Omega_2^2=\frac{2dzd\bar{z}}{(1+z\bar{z})^2}$. Now we perform the following transformations in the (say) $+$ side of the shell placed at $U=\epsilon$ (with $\epsilon$ small) 
\begin{equation}
z\to z+ f(z); \bar{z}\to\bar{z}+ \bar{f}(\bar{z});v\to v(1-\tilde{\Omega}(z,\bar{z}) ) -\frac{M^2}{\epsilon^2}\tilde{\Omega}(z,\bar{z}) )+\frac{2M^2}{\epsilon}\tilde{\Omega}(z,\bar{z}) )+\mathcal{O}(\epsilon),
\end{equation}
where $f(z)$ and $\bar{f}(\bar{z})$ are holomorphic and anti-holomorphic functions. These transformations will induce an infinitesimal conformal transformations on the unit two-sphere satisfying the equation (\ref{CKE}). The conformal factor $\tilde{\Omega}(z,\bar{z})$ is expressed in terms of $z$,$\bar{z}$, $f(z)$ and $\bar{f}(\bar{z})$. As long as $\epsilon\neq0$, the transformation is valid and if one sets $\epsilon=0$ then one must also set $\tilde{\Omega}(z,\bar{z})$ equal to zero indicating at the horizon these transformations do not exist. Therefore, the shell placed near the horizon of ERN black hole gives rise superrotation-like soldering transformations \cite{Bhattacharjee}. The intrinsic properties of this shell can be obtained by similar manner as described earlier. 

\section{Memory effect: Extreme RN black hole}\label{sec4}
In this section, we shall study the memory effect for ERN black hole on timelike geodesics. To study memory effect, one needs to calculate geometric quantities and their derivatives at the horizon. We also need to extend components of geodesic tangents and  deviation vectors in a near vicinity of the horizon shell orthogonally. A natural choice for this should be a Kruskal-like coordinate system that is regular at the event horizon. In \cite{CARTER1966423}, a maximal analytic extension of ERN black hole was constructed by overlapping two sets of double null coordinates. The construction is useful for examining global features of the spacetime but due to presence of trigonometric functions, it provides less analytical control at the event horizon (for example, the metric is not $C^1$ at the event horizon). In Kruskal type coordinates, one uses an exponential mapping from usual advanced or retarded null coordinates which provides a better analytical control. However, for extreme case since the horizon is degenerate (surface gravity $\kappa=0$) the Kruskal coordinates (eg. $u=-{1\over \kappa}\ln({-U}))$ are constant for any value of advanced or retarded coordinates. This can be remedied using a slightly modified version of Kruskal coordinates as presented in \cite{Liberati:2000sq}.

 Therefore, we shall adopt a Kruskal extension that unambiguously places the shell at $U=0$, and also better suited for memory effect analysis. 
 
Recall $u=t-r_*;\,v=t+r_*$ and the tortoise coordinate is given by
\begin{align}\label{pole}
r_{*} = r+2M\Big(ln(r-M)-\frac{M}{2}\frac{1}{(r-M)}\Big)+constant \,.
\end{align}
Although $r_*$ can be made continuous at the event horizon ($r=M$)  by taking the $Q^2\to M^2$ limit from the $r_*$ of non-extreme case (here $Q$ is the charge parameter of RN black hole) \cite{Liberati:2000sq}, but the Kruskal transformations become divergent exactly at $r=M$ as they are related by
\[u=-{1\over \kappa}\ln({-U}); \,v={1\over \kappa}\ln({V}),\]
and $\kappa=0$ at the horizon. Therefore the following transformations for the metric (\ref{RN}) are to be made \cite{Liberati:2000sq}
\begin{align}
u = -\psi(-U) \hspace*{5mm} ; \hspace*{5mm} v = \psi(V),
\end{align}
where we consider $\psi(V)$ to be of the form\footnote{In general for any $\xi$, $\psi(\xi)=4M\Big(ln \xi-\frac{M}{2\xi}\Big)$.}
\begin{align}
\psi(V)= 4M\Big(ln V-\frac{M}{2V}\Big).
\end{align}
Near the horizon
\begin{align}
r_{*}\sim\frac{1}{2}\psi(r-M).
\end{align}
Under this assumption, metric is given by,
\begin{align}
ds^{2} = -\frac{(r-M)^{2}}{r^{2}}\psi(-U)^{'}\psi(V)^{'} dU dV +r^{2}d\Omega_{2}^{2},
\end{align}
where prime denotes the derivative of function $\psi(-U)$ with respect to $U$ and derivative of function $\psi(V)$ with respect to $V$. The transformations are not well defined if the metric is degenerate on the horizon. However, we construct the asymptotic form of the metric as one can have, in the asymptotic limit, $t\sim r_{*}$, and $u\sim -2r_{*} \sim -\psi(r-M)$. Therefore, the inverse transformation is
\begin{align}
U = -\psi^{-1}(-u)\sim -\psi^{-}(\psi(r-M)) = -(r-M),
\end{align}
together with
\begin{align}
\psi(-U)^{'} \sim \frac{4M}{r-M}+\frac{2M^{2}}{(r-M)^{2}}.
\end{align}
$\psi(V)^{'}$ is regular as it is finite and non-zero everywhere. Thus we have a set of Kruskal coordinates that are well defined on the horizon. The metric is written as \cite{Liberati:2000sq}
\begin{align}
ds^{2} = -\frac{2M^{2}}{r^{2}}\psi(V)^{'} dU dV+r^{2}(U)d\Omega_{2}^{2}.\label{lib}
\end{align}
We would consider this metric and study the off-shell extension of the transformations and memory effect. The construction follows the one considered in \cite{PhysRevD.100.084010}. We must mention here, for an extreme-Kerr metric, similar kind of  $(U,V)$ coordinates can be obtained as the structure of $r_*$ is almost identical to (\ref{pole}). However, for obtaining shell's intrinsic properties and to study memory effect, one needs to find the extension of soldering freedom off the shell. For a full 4-dimensional rotating metric, this is not an easy task. 
\subsection{Off-Horizon Shell Extension of Soldering Transformations}
To determine the stress-tensor supported on the horizon shell, we extend the soldering transformations off the horizon shell by making an expansion to linear order in $U$  to one side of the shell \cite{Blau:2015nee}. We shall call this as ``off-shell'' extension of soldering transformations. Of course, the soldering procedure should be consistent with the junction conditions so that the metric remains continuous across the horizon shell. We also find the generators for off-shell soldering transformations and determine the exact off-shell coordinate transformations to the linear order of $U$. More detailed description can be found in \cite{Blau:2015nee}.

We extend the soldering transformations off the horizon shell in $\mathcal{M}_{+}$ side for extreme RN black hole.  As $U=-(r-M)$, and $r=M$ on the horizon, $U=0$ on the horizon shell. The off-shell soldering transformations to the linear order in $U$ are given by
\begin{align}\label{ern-offc}
U_{+} = U C(V,x^{A}) \hspace*{4mm} ; \hspace*{4mm} V_{+} = F(V,x^{A})+UA(V,x^{A}) \hspace*{4mm} ; \hspace*{4mm} x_{+}^{A}=x^{A}+UB^{A}(V,x^{A}),
\end{align}
where $x^{A}$ denote $(\theta,\phi)$. We first need to determine the functions $A(V,x^{A})$ $C(V,x^{A})$ and $B(V,x^{A})$. For this, we take the transformed metric and compare $g_{U\alpha}$ components with the non-transformed metric. We determine
\begin{align}
C =& \frac{\partial_{V}\psi (V)}{\partial_{V}\psi (F)}\\
A =& \frac{M^{2}}{2}\frac{F_{V}}{\partial_{V}\psi(V)}\sigma_{AB} B^{A} B^{B}\\
B^{A} =& \partial_{V}\psi(V)\frac{1}{M^{2}F_{V}}\sigma^{AB} F_{B},
\end{align}
where $\sigma_{AB}$ is unit 2-sphere metric. Now, we specialize our calculations for BMS case i.e. for $V\rightarrow V+T(\theta,\phi)$. We can explicitly write the metric components linear in $U$. 
\begin{align}
Ug_{ab}^{(1)+}dx^{a}dx^{b} =& 2M^{2}U\Big[\Big(-\frac{\psi(V)^{''}}{M^{2}}+\frac{\psi(V)^{'}}{M^{2}\psi(T)^{'}}\psi(T)^{''} \Big)dV^{2} \nonumber
 \\
 & +\frac{\psi(V)^{'}}{M^{2}}\Big(\frac{\partial_{A}\psi(T)^{'}}{\psi(T)^{'}}+T_{A}\frac{\psi(T)^{''}}{\psi(T)^{'}} \Big)dVdx^{A} \nonumber \\ 
& +\frac{\psi(V)^{'}}{M^{2}}\Big(T_{AB}+T_{B}\frac{\partial_{A}\psi(T)^{'}}{\psi(T)^{'}}-\frac{M\sigma_{AB}}{\psi(T)^{'}}\Big)dx^{A}dx^{B} \nonumber\\
& -\frac{2\psi(V)^{'}}{M^{2}}T_{\phi}\cot\theta d\theta d\phi+B^{\theta}\sin\theta\cos\theta d\phi^{2}\Big],
\end{align}
where $\psi(T)^{'}$ denotes the derivative of function $\psi(T)$ with respect to $T$ and $\psi(V)^{'}$ is derivative of function $\psi(V)$ with respect to $V$. For $\mathcal{M}_{-}$ manifold, the metric is
\begin{align}
Ug_{ab}^{(1)-}dx^{a}dx^{b} = -2MU\sigma_{AB}dx^{A}dx^{B}
\end{align}
 Recall the normalization conditions $n^{\mu}n_{\mu}=0$ and $n^{\mu}N_{\mu}=-1$.  The auxiliary normal is chosen to be $N^{\mu}=(1,0,0,0)$.
Now we can extract all components of $\gamma_{ab}$ using
(\ref{gamma from Kab}). The  components of $\gamma_{ab}$ are given by
\begin{align}
\gamma_{VV} =& 2\Big(-\psi(V)^{''}+\frac{\psi(V)^{'}}{\psi(T)^{'}}\psi(T)^{''}\Big) \\
\gamma_{VA} =& 2\frac{\psi(V)^{'}\psi(T)^{''}}{\psi(T)^{'}}T_{A}  \\
\gamma_{\theta\theta} =& 2\psi(V)^{'}\Big(T_{\theta\theta}+\frac{T_{\theta}^{2}\psi(T)^{''}}{\psi(T)^{'}}-\frac{M}{\psi(T)^{'}}+\frac{M}{\psi(V)^{'}}\Big) \label{gtt}\\
\gamma_{\phi\phi} =& 2\psi(V)^{'}\Big(T_{\phi\phi}+T_{\phi}^{2}\frac{\psi(T)^{''}}{\psi(T)^{'}}-M\frac{\sin^{2}\theta}{\psi(T)^{'}}+T_{\theta}\sin\theta\cos\theta+M\frac{\sin^{2}\theta}{\psi(V){'}}\Big) \label{gpp}\\
\gamma_{\theta\phi} =& \gamma_{\phi\theta} =  2\psi(V)^{'}\Big(\frac{T_{\theta}T_{\phi}}{\psi(T)^{'}}\psi(T)^{''}+T_{\theta\phi}-T_{\phi}\cot\theta\Big)	
\label{gm2}\end{align}
Next, we  directly compute the transverse traceless components $\hat{\gamma}_{ab}$
\begin{align}
\hat{\gamma}_{\theta\phi} =& \hat{\gamma}_{\phi\theta} = \gamma_{\theta\phi} \label{gamma1} \\
\hat{\gamma}_{\theta\theta} =& \frac{1}{2}\Big(\gamma_{\theta\theta}-\frac{\gamma_{\phi\phi}}{\sin^{2}\theta}\Big) \label{gamma2} \\
\hat{\gamma}_{\phi\phi} =& \frac{1}{2}\Big(\gamma_{\phi\phi}-\gamma_{\theta\theta}\sin^{2}\theta\Big). \label{gamma3}
\end{align}\label{hat}
Other components can also be calculated in the same way. We can estimate surface energy density in the following manner
\begin{align}
\mu =& -\frac{1}{16\pi M^{2}}\left(\gamma_{\theta\theta}+\frac{1}{\sin^{2}\theta}\gamma_{\phi\phi}\right).
\end{align}
Therefore
\begin{align}
\mu = -\frac{\psi(V)^{'}}{8\pi M^{2}}\Big(\bigtriangleup^{(2)}T(\theta,\phi)-2M\Big(\frac{1}{\psi(T)^{'}}-\frac{1}{\psi(V)^{'}}\Big)+\Big(T_{\theta}^{2}+\frac{T_{\phi}^{2}}{\sin^{2}\theta}\Big)\frac{\psi(T)^{''}}{\psi(T)^{'}}\Big).
\end{align}
The surface current and surface pressure have the following forms
\begin{align}
    J^{A} =& \frac{1}{8\pi M^{2}}\sigma^{AB} \Big(T_{B}\frac{\psi(T)^{''}}{\psi(T)^{'}}\Big) \label{J}\\
    p =& -\frac{1}{8\pi} \frac{1}{(\psi(V)^{'})^{2}} \Big(-\psi(V)^{''}+\frac{\psi(V)^{'}}{\psi(T)^{'}}\psi(T)^{''}\Big),
\end{align}
where $\sigma^{AB}$ is the inverse of the unit 2-sphere metric. Here, unlike the EF shell, we get non-vanishing current and pressure. 
\subsection{Memory Effect for Timelike Geodesics}\label{mtimelike}
Let us now consider two timelike geodesics crossing the horizon shell supporting IGW. We determine the change in the deviation vector between two nearby geodesics passing the horizon shell. We work in a local coordinate system in which the metric is continuous but its first derivative is discontinuous across the null surface. Effect of IGW on geodesics have been also studied in different setups  
 \cite{Balasin_1997, R_stein, *stein, p1, *p2, *p3, *p4, *p5, *p6, *p7, PhysRevD.81.124035}. Here we follow the set up depicted in \cite{PhysRevD.100.084010, C_Barrabes, doi:10.1142/S0218271801001098}. 
 
Let us consider $T^{\mu}$ be a unit timelike vector field with
\begin{align}\label{ortho}
g_{\mu\nu}T^{\mu}T^{\nu} = -1.
\end{align}
Integral curve of $T^{\mu}$ passes through the null shell situated at $\Sigma$. Consider the deviation vector between two nearby timelike geodesics is $X^{\mu}$ with $g_{\mu\nu}T^{\mu}X^{\nu}=0$. The geodesic equation for $X^{\mu}$ is
\begin{align}
\ddot{X}^{\mu} = -R^{\mu}{}_{\rho\sigma\delta}T^{\rho}X^{\sigma}T^{\delta}.
\end{align}
We assume the jump in the derivatives of $T^{\mu}$ and $X^{\mu}$ are proportional to the normal $n^{\mu}$,
\begin{align}\label{WP}
    [\partial_{\beta}T^{\mu}]=P^{\mu}n_{\beta};\,\,[\partial_{\beta}X^{\mu}]=W^{\mu}n_{\beta},
\end{align}
for some $P^{\mu}$ and $W^{\mu}$ defined on $\Sigma$. Integration of the second condition in (\ref{WP}) expresses $X^{\mu}$ off the shell to linear order in $U$. Further, we set up a triad $\{E_a\}$ by parallel transporting $\{e_a\}$s (defined in (\ref{sec2}) ) on $\Sigma$, along the timelike geodesics generated by $T^{\mu}$. Next, $X^\mu$ can be decomposed as
\begin{equation}\label{x-decomposed}
    X^{\mu}=X_0T_{(0)}^{\mu}+X_0^{a}e^{\mu}_a,
\end{equation}
where $X_0, X^a_0$ are some functions (evaluated at $\Sigma$), and the vector fields evaluated at $\Sigma$ are denoted by a subscript $(0)$. With the help of (\ref{ortho}),(\ref{WP}),(\ref{x-decomposed}) and introducing a set of basis vectors $\{E^{\mu}_a\}$ for three vector fields $\{E_a\}$, one can express $X^{\mu}$, for small $U>0$, as \cite{doi:10.1142/S0218271801001098, C_Barrabes}
\begin{equation} \label{X-Uexp}
X_{a} = \Big(\tilde{g}_{ab}+\frac{1}{2}\,U\, \gamma_{ab}\Big)X^{b}_{(0)}+U\,V_{(0)a}^{-}.
\end{equation}
where $V_{(0)a}^{-}=\frac{dX_{a}^{-}}{dU} \big |_{U=0}$ and $\tilde{g}_{ab}$  given by,
\begin{equation} \label{gtilde}
\tilde{g}_{ab} = g_{ab} + (T_{(0)\mu}e^{\mu}_{a})(T_{(0)\nu}e^{\nu}_{b}).
\end{equation}

Next, we decompose $\gamma_{ab}$ in (\ref{X-Uexp}) into transverse and traceless part in order to see the effect of impulsive wave and stress-energy of the shell separately.
\begin{align}
\gamma_{ab} = \hat{\gamma}_{ab} + \bar{\gamma}_{ab},
\end{align} 
where\begin{align}
\bar{\gamma}_{ab} = 16\pi\Big(g_{ac} S^{cd}N_{d}N_{b}+g_{bc}S^{cd}N_{d}N_{a}-\frac{1}{2}g_{cd}S^{cd}N_{a}N_{b}-\frac{1}{2}g_{ab}S^{cd}N_{c}N_{d}\Big).
\end{align}
 From here, choosing $e^{\mu}_V=n^{\mu}$ and $N^{\mu}=T^{\mu}_{(0)}$ \cite{doi:10.1142/S0218271801001098}, we obtain
\begin{align}
\bar{\gamma}_{VB} = 16\pi g_{BC}S^{VC} \hspace*{4mm} ; \hspace*{4mm} \bar{\gamma}_{AB} = -8\pi S^{VV}g_{AB},\label{GammaAB}
\end{align}
where $\bar{\gamma}_{VB}$ is symmetric in lower indices. It is to be noted, choosing the transverse vector $N^{\mu}=T^{\mu}_{(0)}$ means $N\cdot N=-1$, which is consistent with the conditions depicted in Eqs. (\ref{c1}), (\ref{c2}). Now we impose a condition on the test particles that initially they reside on 2-dimensional surface. Thus we set $X_{(0)V}=V^{-}_{(0)V}=0$, where $X_{(0)\mu}$ are components of deviation vector before the passage of the IGW at the horizon, and $V^{-}_{(0)a}=\frac{dX^{-}_{a}}{dU}\Big\vert_{U=0}$. Finally using (\ref{GammaAB}), in this special frame the deviation vectors become
\begin{align}\label{X_A}
X_{V} =& \frac{1}{2}U\bar{\gamma}_{VB} X^{B}_{(0)} = 8\pi U g_{BC}S^{VC}X^{B}_{(0)} \\ 
X_{A} =& X_{A(0)}+\frac{U}{2}\gamma_{AB}X^{B}_{(0)}+UV^{-}_{(0)A}.
\label{X_B}\end{align}
The term involving $\gamma_{AB}$ represents the distortion effect of the wave on the test particle.  Note that, if the surface current is non-zero i.e. $S^{VC}\neq 0$ (see (\ref{ST}),  then $X_{V}\neq 0$. This means the test particle will no longer reside on 2-dimensional surface. It gets displaced off the surface. We also observe that in the case under study, we have a non-zero current or non-zero $S^{VC}$. Hence, the effect of passage of impulsive wave is to deflect the particles off the 2-d surface. The spatial components can also be obtained from (\ref{X_B}). The expression for the $X_{\theta}$ component of the deviation vector
\begin{align}\label{dS}
X_{\theta} = X_{\theta(0)}+\frac{U}{2}\Big(\gamma_{\theta\theta}X^{\theta}_{(0)}+\gamma_{\theta\phi}X^{\phi}_{(0)}\Big) + UV^{-}_{(0)\theta}.
\end{align}
The $X_{\phi}$ component can also be recovered in the same way. One can now replace the requisite components of $\hat{\gamma}_{AB}$ from (\ref{gamma1}-\ref{gamma3}) and $\bar{\gamma}_{AB}$  from (\ref{GammaAB}) 
into (\ref{dS}) to explicitly see the effect of IGW and matter part on the deviation vector.

Suppose the surface current is zero, i.e. $J^A=0$, then (\ref{X_A}) implies $X_V=0.$ Replacing the second expression of (\ref{GammaAB}) into  (\ref{X_B}) yields
\begin{align}\label{dwave}
    X_{A} =(1-4\pi U S^{VV})(\delta_{AB}+\frac{1}{2}U\hat{\gamma}_{AB})X^B_{(0)} .
\end{align}
From (\ref{J}), it is apparent that the anisotropic-stress or surface current of the shell becomes zero for a constant $T$, or constant shifts of $V$. In this case, from (\ref{gtt}), (\ref{gpp}), (\ref{gamma2}), and (\ref{gamma3}) it can be seen that the pure gravitational wave components are identically zero.  So for a shell without surface current we don't have any effect of gravity wave on the test particles as opposed to the case studied for a Schwarzschild black hole in \cite{PhysRevD.100.084010}.  Therefore it is impossible to have a relative displacement of the test particles confined purely in the 2-d plane for IGW supported at an ERN horizon shell. The particles will always be displaced from their initial plane. Clearly, the deviations between two timelike geodesics are determined in terms of supertranslation parameter $T(\theta,\phi)$ contained in $\gamma_{AB}$ or the stress-tensor intrinsic to the shell.  We can integrate the expressions of the  components of deviation vectors and obtain the shift with respect to the parameter of the geodesics. This gives us ``displacement memory effect''. For physical shells having $S^{VV}>0$, we can also see there is a diminishing effect for the transverse components of the deviation vectors carrying the BMS-like memories. 
\section{B-tensor and Null geodesics}\label{mNull}
In this section, we shall study the effect of IGW on null geodesics passing orthogonally through the horizon shell. We would estimate the jump in optical parameters such as shear, expansion. Throughout this study, we  assume a continuous coordinate system is installed covering both sides of the null hypersurface or horizon, and the components of the geodesic tangent vector crossing the horizon are continuous in this coordinate system.  Let us consider a null congruence whose tangent vector is denoted by $\CN$ and it is orthogonally crossing the hypersurface $\Sigma$ supporting IGW \footnote{The chosen congruence obeys hypersurface orthogonality \cite{OLoughlin:2018ebk}. }. The normalization condition with the normal to the horizon will be as usual $n\cdot \CN=-1,n\cdot n=0$ and $\CN\cdot e=0$. The effect of null geodesic crossing the surface containing IGW from (say) ``-'' side to the ``+'' side is to apply a coordinate transformation from past coordinates to a continuous coordinate system $(x^{\mu})$ at $\Sigma$. This coordinate transformation then serves as the initial condition to develop it to the future or ``+'' side of the shell. This construction is similar as described in (\ref{sec2}), only for assigning the past and future of the shell unambiguously, the off-the shell coordinate transformations are done in the ``-'' side. This has practically no effect on the geometric constructions outlined in previous sections. 

The major object of interest in our study is the failure tensor or the $B$-tensor with respect to the vector $\CN_0$ projected onto the hypersurface $\Sigma$.
\begin{equation}\label{B-def}
   B_{AB}=B_{\alpha\beta}e^{\alpha}_Ae^{\beta}_B= e^{\alpha}_Ae^{\beta}_B\nabla_{\alpha}\CN_{0\beta}.
\end{equation}
The expansion $\Theta$ and shear $\Sigma$ are expressed in terms of $B$-tensor in the familiar way-

\[\Theta= \sigma^{AB}B_{AB};\,\,\Sigma_{AB}=B_{(AB)}-\dfrac{\Theta}{2}\sigma_{AB},\]
where $\sigma^{AB}$ is the inverse of the metric induced at the spatial section of the hypersurface or the shell. Next, we determine the off-shell B-tensor by pulling back the B-tensor at the shell to an infinitesimal distance away from the shell using \cite{OLoughlin:2018ebk, PhysRevD.100.084010}

\begin{equation}\label{coord} x^{\mu}=x^{\mu}_0+U\CN_{0}^{\mu}(x^{\mu}).\end{equation}
Here, the coordinates on the hypersurface are designated as $x_0^{\mu}$ and related to the continuous coordinates $x^{\mu}$ via above relation. The vector $\CN_0$ provides the initial condition for obtaining the coordinates $x^{\mu}$ off the shell.
\begin{align}\label{Jcbn}
\tilde{B}_{AB}(x^{\mu}) = \frac{\partial x_{0}^{M}}{\partial x^{A}} \frac{\partial x_{0}^{N}}{\partial x^{B}}B_{MN}(x^{\mu}_{0}).
\end{align}
   We expect a non-vanishing change in optical parameters which depicts a kind of memory effect on the geodesics.  Since the B-tensor and quantities derived from it encode the effect of stress-tensor and IGW supported at the shell, this covariant version of memory is regarded as \emph{B}-memory.

\subsection{B-memory Effect for Null Geodesics}
We consider extreme RN black hole in Kruskal coordinates given by (\ref{lib}). The tangent vector in past side to the null congruence is $\CN_{-} = \lambda \partial_{U}$. Components of congruence $\CN_{0}$ on hypersurface in continuous coordinates are obtained using (\ref{coord}) and  calculating the inverse Jacobian of coordinate transformation  (see Appendix \ref{apnB})
\begin{align}\label{N_0}
\CN_{0}^{\alpha} = \Big(\frac{\partial x^{\beta}_-}{\partial x^{\alpha}}\Big)^{-1} \CN_{-}^{\beta}\Big\vert_{\Sigma}.
\end{align}
We first find the tangent vector at the hypersurface
\begin{align}
\CN_{0} = \lambda\Big(\frac{F_{V} \psi (F)'}{\psi (V)'}\partial_{U}+\frac{\psi (F)'}{2 M^2 F_{V}}  \left(F_{\theta}^2+ \frac{F_{\phi}^2}{\sin^{2}\theta}\right) \partial_{V}-\frac{F_{\theta} \psi (F)'}{M^2}\partial_{\theta}-\frac{F_{\phi} \psi (F)'}{M^2 \sin^{2}\theta}\partial_{\phi}\Big)\Big\vert_{\Sigma}.
\end{align}
From $U$-component of $\CN_{0}$, we determine
\begin{align}
\lambda = \frac{\psi(V)^{'}}{\psi(F)^{'}F_{V}}.
\end{align}
Using eqn. (V.2) and $\CN_{0}$ expression, we get
\begin{align}
B_{AB} = 2\frac{\psi(V)^{'}}{F_{V}^{2}}F_{A}F_{BV}-\frac{\psi(V)^{'}}{F_{V}^{3}}F_{A}F_{B}F_{VV}-\frac{\psi(V)^{'}}{F_{V}}F_{AB}+\frac{F_{A}F_{B}}{F_{V}^{2}}\psi(V)^{''}-\Gamma^{\delta}_{AB}\CN_{0\delta}.
\end{align}
We will consider now the expressions of $\Theta$ and $\Sigma_{AB}$ evaluated with respect to $\CN_0$ and $\CN_-$ to find the jumps in these optical tensors. If we specialize our case for BMS supertranslation, the non-vanishing change in expansion and shear on the shell (at $U=0$) are
\begin{align}
[\Theta] =& \frac{1}{M^{2}} \Big(-\psi(V)^{'} \bigtriangleup^{(2)}T(\theta,\phi) +\psi(V)^{''}\sigma^{AB}T_{A}T_{B}\Big)\\
[\Sigma_{\theta\theta}] =& \frac{\psi(V)^{'}}{2}\Big(-T_{\theta\theta}+\frac{T_{\phi\phi}}{\sin^{2}\theta}+T_{\theta}\cot\theta\Big)+\frac{\psi(V)^{''}}{2}\Big(T_{\theta}^{2}-\frac{T_{\phi}^{2}}{\sin^{2}\theta}\Big) \\ 
[\Sigma_{\phi\phi}] =& \frac{\psi(V)^{'}}{2}\Big(-T_{\phi\phi}+T_{\theta\theta}\sin^{2}\theta-\frac{T_{\theta}\sin 2\theta}{2}\Big)+\frac{\psi(V)^{''}}{2}\Big(T_{\phi}^{2}-T_{\theta}^{2}\sin^{2}\theta\Big) \\
[\Sigma_{\theta\phi}] =& -\psi(V)^{'}T_{\theta\phi}+T_{\theta}T_{\phi}\psi(V)''+\psi(V)^{'}T_{\phi}\cot\theta
\end{align}
where $\sigma^{AB}$ is inverse of unit 2-sphere metric. Here, we have non-vanishing jumps for expansion and shear comprising BMS parameter $T(\theta,\phi)$ and its derivatives. The jump in the expansion and shear are  determined by the shell stress-energy tensor and a combination of IGW and stress-tensor respectively. Here the term \emph{B}-memory is used to recognise the fact, the interaction of the shell with test detectors give rise changes in optical tensors, and these changes are expressed via BMS soldering parameters. This is similar to the BMS memory effect that one obtains in the far region of asymptotically flat spacetimes.

\subsection{B-memory Effect in Off-Shell Extension of Null Congruence}
Now we compute the off-shell extended $B$-tensor for the soldering transformation of supertranslation type. To find the Jacobians of the transformation used in (\ref{Jcbn}) we write \cite{OLoughlin:2018ebk, PhysRevD.100.084010}
\begin{align*}
P^M_A=\frac{\partial x_0^M}{\partial x^A}=(W^{-1})^M_A,\end{align*}
where $W^M_A=(\delta^{M}_{A}-U\frac{\partial \CN^{M}}{\partial x^{A}})$, and compute the inverse of the $W$ matrix as
\begin{align}
 \frac{1}{det(W)}\left(
\begin{array}{cc}
 1+\frac{U\psi(V)^{'}}{M^{2}}\frac{T_{\phi\phi}}{\sin^{2}\theta} & -\frac{U\psi(V)^{'}}{M^{2}}T_{\theta\phi} \\
 -\frac{U\psi(V)^{'}}{M^{2}}\Big(\frac{T_{\theta\phi}}{\sin^{2}\theta}-\frac{2\cos\theta T_{\phi}}{\sin^{3}\theta}\Big) &  1+\frac{U\psi(V)^{'}}{M^{2}}T_{\theta\theta}
\end{array}
\right).
\label{Jacobian4}
\end{align}
Where 
\begin{align*}
det(W) =\Big(1+\frac{U\psi(V)^{'}}{M^{2}}\frac{T_{\phi\phi}}{\sin^{2}\theta}\Big)\Big(1+\frac{U\psi(V)^{'}}{M^{2}}T_{\theta\theta}\Big) -\frac{U^{2}\psi(V)^{'2}}{M^{4}}T_{\theta\phi}\Big(\frac{T_{\theta\phi}}{\sin^{2}\theta}-\frac{2\cos\theta T_{\phi}}{\sin^{3}\theta}\Big).
\end{align*}
Here we also used the fact, $\CN(x^{\mu})=\CN_0(x_0^{a}(x^{\mu}))$\footnote{This follows from the hypersurface orthogonality of the congruence.}. The full expression of (\ref{Jcbn}) is quite huge to be written here as it will be multiplication of two $W^{-1}$ matrices. However, we  display the shortest component of the $B$-tensor up to linear order in $U$ here.
\begin{align}
B_{\theta\theta} =& (\psi ''(V) T_{\theta}^{2}-\psi '(V) T_{\theta\theta}-M)+\frac{2U}{M^{2}}\Big( -\cot (\theta ) \psi '(V)^2 T_{\phi} T_{\theta\phi}-\psi '(V) \psi ''(V) T_{\phi} T_{\theta} T_{\theta\phi}+ \nonumber \\
& M \psi '(V) T_{\theta\theta}+\psi '(V)^2 T_{\theta\phi}^2+\psi '(V)^2 T_{\theta\theta}^2-\psi '(V) \psi ''(V) T_{\theta\theta} T_{\theta}^2 \Big)\,+ {\cal{O}}(U^2).
\label{B-tt}\end{align}
If one sets $U=0$, the off-shell $B$-tensor reduces to on-shell $B$-tensor which we have already computed in the previous subsection. it is clear from (\ref{B-tt}) that the off-shell $B$-tensor also suffers jump ($U$ dependent) across the null shell and the jump is parametrized by BMS parameters. Thus, B-memory effect is again visible in the off-shell extension of the soldering transformation. An alternative approach to see the change in $B$-tensor for the null geodesics crossing the shell could be to consider the Lie derivative of $B$-tensor with respect to the vector field $\CN_{0}$. This would also give rise a B-memory effect for null geodesics.   

\section{ Discussion}
The motivation of this work is to find the memory effect of IGW supported at a horizon shell of extreme black holes for timelike and null geodesics (or test detectors) crossing the null shell. Although the memory corresponding to BMS type symmetries discussed here is quite distinct from the memory being studied in the far region or asymptotic infinities of black holes, but this study may serve as a model for determining the effect of impulsive gravity waves (together with some thin layer of null matter like neutrino fluid), generated during violent astrophysical phenomena, on test particles. The appearance of BMS-like symmetries at the horizon or at any null surface situated at a finite distance of a spacetime, provides an intriguing possibility to investigate direct evidence of those symmetries. Our attempt here is to provide a theoretical model that can capture such symmetries. It would be interesting to see how our considerations can be related to recent studies of BMS symmetries on null hypersurfaces and local horizons \cite{Ashtekar,Chandrasekaran}. 

Although the mathematical frameworks used in this note are already applied to study BMS-memory effect for non-extreme black holes in \cite{PhysRevD.100.084010}, but several new features have also been obtained in this study for the extreme black holes. The nature of horizon-shells containing BMS memories for ERN black hole has non-vanishing surface current as opposed to the case of non-extreme (Schwarzschild) black holes \cite{PhysRevD.100.084010}. In fact one can't have a physical shell carrying BMS like charges with a vanishing surface current for ERN case. This feature changes the way test (timelike) particles get deflected from their initial position after passage of IGW. There are also other novel features of this study which we summarize below in more detail.

\begin{itemize}
\item[1.] We have shown how BMS supertranslation-like symmetry arises as soldering freedom for a horizon shell in ERN black hole. We started with estimating the intrinsic properties of the null shell for ERN black hole. We also study the BMS type soldering freedom for extreme BTZ case. The detailed study for BTZ black hole is shown in Appendix (\ref{apnA}).  We observe, for both the cases, there is no possibility of a shell where IGW  and the matter supported on the shell get decoupled. Therefore these horizon shells do not support pure IGW without matter. This is same as the case of non-extreme black holes.
\item [2.]  We also discussed how conformal symmetries may arise as soldering freedom when we place a null shell infinitesimally close to the event horizon of an ERN black hole.  We have related this with the Penrose's cut-paste construction. For an ERN black hole, introducing a double null coordinate system, we have demonstrated the appearance of superrotation-like symmetry near (but not on the) horizon. 

\item[3.] Next, we performed the off-shell extension of soldering transformations for ERN black holes. We used a Kruskal-like double null metric that is regular at the horizon and obtained the off-shell extension to the linear order of U. We then computed shell-intrinsic properties. The Kruskal shell for ERN has non-vanishing pressure and surface current. This is in contrast to the Kruskal shell in Schwarzschild black hole where both of these  intrinsic quantities vanish. Thereafter, we obtained the components of deviation vectors in terms of BMS-like parameters that are present in the shell's stress-energy tensor. We show, the test particles initially at rest get displaced from their initial plane after they interact with the shell supporting IGW and null mater. This corresponds to the memory effect for timelike geodesics crossing the null shell. We have  also  shown there can not be a deflection that will {\emph keep} the test particles on the initial 2-d surface (codimension 1 surface  of the shell) and only induce a relative displacement between them, as was seen for a non-extreme black hole \cite{PhysRevD.100.084010}.

\item[4.] Further, we computed the memory effect for null geodesics passing orthogonally through the horizon shell placed in ERN spacetime. We observed, there is a non-vanishing change in optical tensors (the expansion and shear) which again shows a type of memory effect (or $B$-memory) for null geodesics crossing orthogonally to the null hypersurface. We also get a finite jump in the expansion and shear, for the congruence at points infinitesimally away from the shell. Also due to non-existence of pure IGW, the jumps in shear can't be attributed to the pure gravitational degree of freedom encoded in IGW as was found for flat space \cite{OLoughlin:2018ebk}.
\end{itemize}

The constructions depicted here in principle can be generalized for Kerr and extreme Kerr spacetimes also. A double null or Kruskal type coordinate system, to study the memory effect, can be obtained very much similar manner as described in (\ref{sec4}), but due to the absence of spherical symmetry, the analytic calculation becomes much more challenging. We would like to consider the rotating metrics in $ 4$-dimensions in future. 

It is known that any spacetime metric can be reduced to a plane wave metric around a point of a null geodesic. This was shown by Penrose and the resulted metric is known as Penrose limit of a spacetime \cite{RevModPhys.37.215}. As Penrose limit produces plane wave spacetimes (pp-wave, plane symmetric, homogeneous waves etc.), it is comparatively much easier to study geodesic deviation vectors and optical tensors in those backgrounds.  These studies can provide some useful theoretical setups that may prove useful in the future detection schemes of memory effect.  Using Penrose limit, the conventional non covariant memory effect for many impulsive gravitational wave and shock wave metrics can be studied \cite{shore_memory_2018}. The near horizon limit of ERN black hole has $AdS_2 \times S^2$ geometry. In static coordinates it reads \cite{Townsend:1997ku, lucietti_horizon_2013} 
\begin{align}
    ds^{2} \simeq -\lambda^{2} dt^{2}+\frac{r_{0}^{2}}{\lambda^{2}}d\lambda^{2}+r_{0}^{2}d\Omega^{2}, \label{penrose limit}
\end{align}
where $\lambda = \frac{r-r_{0}}{r_{0}} << 1$ and $r_{0}=M$ is the horizon of the black hole. Penrose limit of this metric produces a symmetric plane wave spacetime \cite{M_Blau}. It will be interesting to study the memory effect in such symmetric plane wave spacetimes. 

Another useful extension of this work is to study different flux-balance laws as depicted in \cite{Compere}. Studying quantum effects in such IGW spacetimes generated in extreme black holes could be another interesting study where semiclassical features of near horizon symmetries may show up \cite{Horta_su_1996}.
\section*{Acknowledgements} 
S. B. is supported by SERB-DST through the Early Career Research award grant no. ECR/2017/002124. S.K. acknowledges illuminating discussions with Arpan Bhattacharyya.  SB acknowledges useful discussions with Matthias Blau. The authors would like to thank the anonymous referee for many useful comments and suggestions that have enhanced the quality of presentation of this work significantly.
\appendix

\section{Soldering Freedom \& Intrinsic Quantities in Extreme BTZ Black Hole}\label{apnA}

In 3-dimensions, BTZ black hole provides a good model where we can study the horizon shells for black holes with rotation. In $4$-dimensions, the analysis becomes quite difficult.  We thus study the soldering freedom for rotating BTZ black holes and try to gain some insight  for the case when black hole possesses angular momentum.

First, we present the intrinsic quantities for extreme BTZ black hole. We solder two rotating extreme BTZ metrics with horizon situated at $r_{0}=r_{\pm}=l\sqrt{\frac{M}{2}}$. $l^2=-\frac{1}{\Lambda}$, and $\Lambda<0$ is the cosmological constant. $M$ is the mass of the BTZ black holes. For $\mathcal{M}_{-}$ manifold, in EF coordinate system, the metric takes the following form
\begin{align}
ds^{2} = -\frac{(r^{2}-r_{0}^{2})^{2}}{r^{2}l^{2}}dv^{2}+2dvdr+r^{2}(d\phi+N^{\phi}dv)^{2}, 
\end{align}
where
\begin{align}
N^{\phi} = -\frac{J}{2r^{2}} \hspace*{3mm} ; \hspace*{3mm}  J = Ml.
\end{align}
In general
\begin{align}
r_{\pm}^{2} = \frac{Ml^{2}}{2}\Big\lbrace 1\pm\Big[1-\Big(\frac{J}{Ml}\Big)^{2}\Big]^{\frac{1}{2}}\Big\rbrace .\label{horizon}
\end{align}
Here, we see when the angular momentum $J$ equals $M l$, $r_{\pm}$ becomes $r_{0}$. Now, considering the supertranslation type transformations as
\begin{align}
v_{+}=v+T(\phi) \hspace*{3mm} ; \hspace*{3mm} r_{+} = r \hspace*{3mm}  ; \hspace*{3mm} \phi_{+}=\phi.
\end{align}
We obtain intrinsic quantities of the shell
\begin{align}
\mu =& -\frac{1}{8\pi}\sigma^{AB}[\kappa_{AB}] = -\frac{1}{8\pi r_{0}^{2}}\Big(\Gamma^{v_{+}}_{\phi\phi} + 2T_{\phi}\Gamma^{v_{+}}_{v_{+}\phi}+T_{\phi}^{2}\Gamma^{v_{+}}_{v_{+}v_{+}}+r_{0} \Big)\\
J^{A} =& \frac{1}{8\pi}\sigma^{AB}[\kappa_{vB}] = \frac{1}{8\pi r_{0}^{2}}\Big(\Gamma^{v_{+}}_{v_{+}\phi}+T_{\phi}\Gamma^{v_{+}}_{v_{+}\phi}\Big)\\
p =& -\frac{1}{8\pi} [\kappa_{vv}] = -\frac{1}{8\pi}(\Gamma^{v_{+}}_{v_{+}v_{+}}-\Gamma^{v}_{vv}) .
\end{align}
We again recover supertranslation in the shell's intrinsic quantities. We also observe, there can't be a shell without matter supporting pure IGW. We skip displaying the long expressions for Christofell symbols as those are not required to comprehend the appearance of BMS type symmetries at the horizon shell. It seems in $3$-dimensions, one can't make any shell that can induce something similar as superrotation-like symmetry.  Due to the periodic identification of angular coordinate, no construction may produce a feasible solution.

\section{Inverse Jacobian}\label{apnB}
 We provide here the inverse Jacobian of the coordinate transformation used in Eq. (\ref{N_0}) for the extreme RN case. The coordinate transformations are given in (\ref{ern-offc}). We just label the $'+'$ coordinates as $'-'$, and compute the components of $\CN_{0}$ on the hypersurface. The Jacobian matrix reads
\begin{align}
\left[\Big(\frac{\partial x_{-}^{\beta}}{\partial x^{\alpha}}\Big)^{-1}\right] =  \left(
\begin{array}{cccc}
 \frac{\psi '(F) F_{V}}{\psi '(V)} & 0 & 0 & 0 \\
 \frac{ \psi '(F) \left(\csc ^2(\theta ) F_{\phi}^{2}+F_{\theta}^{2}\right)}{2 M^2 F_{V}} & \frac{1}{F_{V}} & -\frac{F_{\theta}}{F_{V}} & -\frac{F_{\phi}}{F_{V}} \\
 -\frac{\psi '(F) F_{\theta}}{M^2} & 0 & 1 & 0 \\
 -\frac{\csc ^2(\theta ) \psi '(F) F_{\phi}}{M^2} & 0 & 0 & 1 \\
\end{array}
\right).
\end{align}


\providecommand{\href}[2]{#2}\begingroup\raggedright

\endgroup
\end{document}